\documentclass{elsart3}
\usepackage{amsmath}
\usepackage{amssymb}
\usepackage{graphicx}% Include figure files
\usepackage{dcolumn}% Align table columns on decimal point
\usepackage{bm}% bold math
\def\be{\begin{eqnarray}}
\def\ee{\end{eqnarray}}
\def\ba{\begin{array}}
\def\ea{\end{array}}
\def\nn{\nonumber}
\begin{document}
\begin{frontmatter}
\title{The current polarization rectification of the integer quantized Hall effect}
\author[l1]{D. Eksi},
\author[l1]{O. Kilicoglu},
\author[l1]{S. Aktas}
and
\author[l2]{A. Siddiki}
\address[l1]{Trakya University, Department of Physics, 22030 Edirne, Turkey}
\address[l2]{Physics Department,  Faculty of Arts and
Sciences, 48170-Kotekli, Mugla, Turkey}
\begin{abstract}
We report on our theoretical investigation considering the widths
of quantized Hall plateaus (QHPs) depending on the density
asymmetry induced by the large current within the out-of-linear
response regime. We solve the Schr\"odinger equation within the
Hartree type mean field approximation using Thomas Fermi Poisson
nonlinear screening theory. We observe that the two dimensional
electron system splits into compressible and incompressible
regions for certain magnetic field intervals, where the Hall
resistance is quantized and the longitudinal resistance vanishes,
if an external current is imposed. We found that the strong
current imposed, induces an asymmetry on the IS width depending
linearly on the current intensity.
\end{abstract}
\end{frontmatter}
%
%Deniz Eksi, Trakya University, Department of Physics, 22030 Edirne, Turkey\\
%Fax: +90 711 6891010, Email: D.eksi@trakya.edu.tr\\
\section{Introduction}
The phenomenon of the integer quantized Hall effect
(IQHE)~\cite{vKlitzing80:494} continuous to hold interest as newer
and newer types of
hetero-junctions~\cite{Heiblum05:abinter,Goldman05:155313,josePHYSE}
are produced. The early attempts to explain the IQHE, like the
bulk~\cite{Laughlin81} or the
edge~\cite{Halperin82:2185,Buettiker86:1761} pictures, considers
electron-electron interactions to be irrelevant and attributes the
effect either to disorder or to the bending of the confinement
potentials, respectively. From theses theories, it is known that
the widths of the QHP depend on the electron density, mobility,
temperature and the amplitude of the applied current. However, the
direction of the applied current is not considered to be
influencing the plateau widths. However, the inclusion of the
(direct) Coulomb interaction
numerically~\cite{Wulf88:4218,siddiki2004} or
analytically~\cite{Chklovskii92:4026} enriches the physics beyond
the single particle pictures. The utilization of the local Ohm's
law~\cite{Guven03:115327} together with the self-consistent
numerical calculations allowed Siddiki and Gerhardts to calculate
the quantized Hall plateaus and also the transition between the
plateaus~\cite{siddiki2004}, within the linear response regime. A
further investigation considering the out of linear response
regime showed that the widths of current carrying egde-states
linearly depend on the current intensity based on the
electron-electron interactions~\cite{denizphyEvelocity}.
\begin{figure}
% A minipage that covers half the page
\centering
\includegraphics[width=7cm]{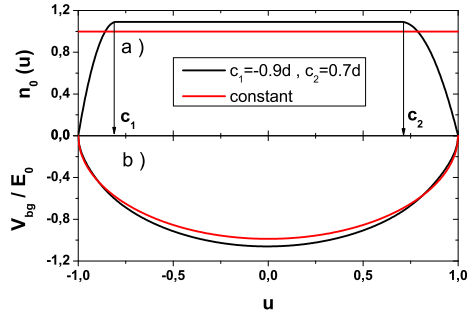}
\caption{ \label{fig:fig1} The cross section of the donor layer
considering (a) for two values of steepness parameters which are
$c_1$ left side and $c_2$ right side parameters of donor
distribution. The red line represents a constant donor
distribution.}
\end{figure}
In this work, we obtain the widths of the QHPs from a model which
is purely based on the electron-electron interactions, supported
by the local Ohm's law ~\cite{siddiki2004}. We solve the
Schr\"odinger and the Poisson equations self-consistently within
the Thomas-Fermi approximation~\cite{Lier94:7757}, which
implicitly assumes that the potential landscape varies slowly on
the quantum mechanical length scales. We start from a homogeneous
donor distribution to calculate the confinement potential (in
fig.~\ref{fig:fig1}a depicted with red line), which we use as an
initial condition for our iterative numerical technique. We then
consider an inhomogeneous distribution of the donors to obtain
different potential slopes at the two edges of the sample (in
fig.~\ref{fig:fig1}a depicted with black line). Background
potentials generated by the donor distributions are shown in
fig.~\ref{fig:fig1}b with the same color code. As we show later,
by doing so we directly change the widths of the incompressible
strips (ISs) resulting from the screening. Within these regions
the backscattering is suppressed, therefore current is confined at
the ISs, hence any effect that influences the widths of the ISs
will effect the current and potential distribution in the sample.
It was shown that, if there exists an IS somewhere in the sample
the system is in a QHP~\cite{Bilayersiddiki06:}. The
self-consistent model, predicts that the widths of the ISs will
also be modified by the imposed current, namely by the
amplitude~\cite{SiddikiEPL:09}. If a DC current is passed in the
+y direction, due to the tilting of the Landau levels, the IS at
the right hand side (RHS) enlarges, whereas, the IS on the left
hand side (LHS) shrinks. Fig.~\ref{fig:fig2}a, depicts such a
situation under current bias. Now if we start with a narrow IS on
the LHS, it is possible to achieve equi-width ISs on both sides,
by applying a certain imposed current, fig.~\ref{fig:fig2}b. As a
result, we conclude that the widths of the QHPs also should depend
on the applied current direction~\cite{SiddikiEPL:09}. To
summarize, by our self-consistent calculations we show that, the
widths of the QHPs also depend on the current direction, which is
in strong contrast to the conventional approaches.

The calculation scheme starts by determining the boundary
conditions to describe the electronic system at hand: First, we
assume a translation invariance in the current direction,
\emph{i.e.} $y-$, hence the electrostatic potential (therefore the
$y$ component of the electric field is also constant, $E_y^0$),
second we consider a lateral confinement in $x$ direction
generated by a donor distribution $n_0(x)$ limited by top-side
gates, which imposes the boundary conditions $V(-d)=V(d)=0$, where
$d$ is the sample width. The analytical solution of the Poisson
equation considering the above boundary conditions reads to the
kernel
\begin{equation}
K(x,x')=\ln\left|\frac{\sqrt{(d^2-x^2)(d^2-x^{'2})}+d^2-x'x}{(x-x')d}\right|
\label{kernel}.
\end{equation}
\begin{figure}
% A minipage that covers half the page
\centering
\includegraphics[width=7cm]{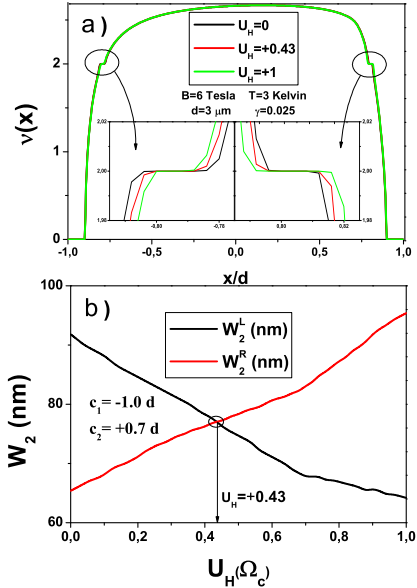}
\caption{ \label{fig:fig2} The electron density as a function of
lateral coordinate normalized with the sample width (a), for three
selected current amplitudes ($U_{\rm H}$). Insets depict the
regions, where incompressible strips reside. It is clearly seen
that the IS at LHS becomes narrower by increasing UH, and opposite
for the RHS. The widths of the ISs as a function of UH, when
applying a positive current one can obtain  equi-width ISs on both
edges, regardless of the donor in homogeneity (b).}
\end{figure}

The confinement potential is obtained by the following integration
for a given $n_0(x)$
\begin{equation}
V_{\rm
bg}(x)=\frac{2e^2}{\bar{\kappa}}\underset{-d}{\overset{+d}\int}dx'n_{
0}(x')K(x,x') \label{conf},
\end{equation}
where $e$ is the electronic charge, $\bar{\kappa}$ an average
dielectric constant and yields to
\begin{equation}
\small{V_{\rm bg}(x)=-E_{\rm bg}^0\sqrt{1-(x/d)^2}
\label{background},\quad  E^0_{\rm bg}=2\pi e^2n_0d/\bar{\kappa},}
\end{equation} given that the donors are homogeneously
distributed. However, as will be discussed later, we also consider
an inhomogeneous donor distribution to create an asymmetric
lateral confinement by considering a donor distribution described
as below\be n_0(x)=
 \left\{\begin{array}{cc}
  -\frac{(u+c_1)^2}{(c_1-1)^2}+1, & -1\leq u<c_1 \\
  1, & c_1\leq u<c_2 \\
  -\frac{(u-c_1)^2}{(c_1-1)^2}+1, & c_2\leq u<1
\end{array}\right\}. \nn \ee By doing so we can controllably break
the lateral confinement symmetry by setting $c_1$ and $c_2$
(almost) arbitrarily. Fig.~\ref{fig:fig1}a presents a situation
considering a homogeneous donor distribution (i.e. $-c_1=c_2=1$)
and also a case where left side is more confining than the right
side. Note that the donor number density is kept constant, that is
the area below the the donor distribution curves are equal. The
resulting confinement potentials are shown in
fig.~\ref{fig:fig1}b, one can readily see that the asymmetric
donor distribution leads a steeper bending on the left hand side
(black line). The corresponding electron distribution in the
absence of magnetic field $B$ and vanishing temperature $T$ is
obtained from \be n_{\rm el}(x)=D_0\Theta(V_{bg}(x)-E_F), \ee
where $D_0$ is a constant that corresponds to the two-dimensional
density of states (DOS) in the absence of an applied $B$ field and
$E_F$ is the Fermi energy fixed by the charge neutrality of the
system. The next step is to calculate the interaction potential
(energy) from \be V_{\rm
H}(x)=\frac{2e^2}{\bar{\kappa}}\underset{-d}{\overset{+d}\int}dx'n_{\rm
 el}(x')K(x,x') \label{hartree}. \ee
At finite temperatures the electron density is calculated from \be
n_{\rm el}(x)=\int dE D(E)f(E,\mu,kT,V(x)), \ee where $D(E)$ is
the relevant DOS, $f(\epsilon)$ is the Fermi occupation function
and $\mu$ is the electrochemical potential. Now by solving the
total potential and the electron distribution iteratively, one can
obtain the electrostatic quantities at equilibrium.

Once these quantities are known it is required to have a
prescription which relates the electron density to the local
conductivities~\cite{siddiki2004} considering a fixed imposed
current $I$, in our work we take this prescription from the
self-consistent Born approximation~\cite{Ando82:437}. At a first
approximation one can neglect the effect of the imposed current on
the electrostatic quantities (namely, the linear response) and the
current distribution can be obtained simply by applying Ohm's law
locally~\cite{Guven03:115327}. The Ohm's law states that the
(local) potential drop is proportional to the local current times
the local resistance (resistivity at 2D, with square
normalization), i.e. we should look for drops at the
self-consistently calculated potential. As an oversimplified
picture, now we relate the screening properties of the electron
gas in the presence of $B$ field with the potential drop. Since
the magnetic field Landau quantizes the system, there are two
possibilities when considering the pinning of the Fermi energy to
the Landau levels: 1) the $E_F$ is equal to one of the Landau
level, \emph{the compressible state}, hence the DOS is high, and
the system behaves like a metal. Therefore, as in all metals, the
potential is constant and no current can flow with in these
regions; 2) the $E_F$ is not equal to the Landau energy the system
is at the \emph{incompressible state} and the self-consistent
potential varies, hence the applied current flows from these
regions. In fig.~\ref{fig:fig2}a the calculated electron densities
(in fact the filling factor, defined as $\nu(x)=2\pi l^2 n_{\rm
el}(x)$, with the magnetic length $l=\sqrt{eB/m}$) are shown
considering an asymmetric donor distribution by setting $c_1=-1$
and $c_2=0.7$. We see that the ISs are formed at both sides where
the potential drops and density is constant considering three
characteristic current biasses, $U_H$, which is measured in units
of cyclotron energy $(\Omega_c$). The ISs are highlighted at the
insets, we see that at higher current densities the left ISs
starts to shrink, whereas the right ISs becomes wider. The IS
width dependency on the current amplitude is shown in
fig.~\ref{fig:fig2}b. It is seen that the donor distribution
asymmetry induced large IS at the left side (red line) starts to
shrink when increasing the bias and its width becomes equal to the
width of the right IS (black line) at $U_H=0.43$. The effect of
large bias current (out of linear response) implies that the
formation of ISs strongly depends on the current amplitude, hence
the QHPs also depends on the the polarization of the current. This
can be seen by considering the slope of the Hall potential, say if
the DC current is positive the Hall potential has an positive
slope or vice versa. Now consider a potential drop at the IS which
has an positive slope, the Hall potential will enlarge the IS on
the right hand side. In the opposite situation the left IS is
enhanced. Therefore depending on the current polarization one of
the ISs will become leaky at a lower $B$, hence the quantized Hall
effect is smeared~\cite{SiddikiEPL:09}. A detailed investigation
of the current polarization on the quantized Hall plateaus is
discussed at Ref.[15].
%\begin{figure}
%\begin{minipage}[b]{0.5\linewidth} % A minipage that covers half the page
%\centering
%\includegraphics[width=8cm]{fig3a.eps}
%\end{minipage}
%\hspace{1.cm} % To get a little bit of space between the figures
%\begin{minipage}[b]{0.5\linewidth}
%\centering
%\includegraphics[width=8cm]{fig3b.eps}
%\end{minipage}
%\caption{\label{fig:fig3} The widths of the incompressible strips
%at filling factor two  at left side of sample as a function of
%sample widths (x- axis) and UH, the applied current in units of
%induced Hall potential (a). Also at the right side of the
%sample(b).}
%\end{figure}

\section{Conclusion}
For the high mobility, narrow and asymmetric samples we predict
that, the large current either enlarges or shrinks the QHPs
depending on whether the asymmetry induced by the current and the
asymmetry caused by the edge profile coincides or not. Based on
our findings, we proposed a sample structure where the effect of
the current induced asymmetry and thereby the rectification of the
QHPs can be controllably measured. As a final remark, we note that
at the edge IQHE regime, due to the competition between the
enhancement of the ISs resulting from the large current and
suppression due to steep potential profile, therefore we expect a
hysteresis like behavior in this regime both depending on the
sweep rate and direction of the B field and current amplitude.

The authors would like to acknowledge the Scientific and Technical
Research Council of Turkey (TUBITAK) for supporting under grant no
109T083 and Mu\~gla University for supporting the "1$^{st}$ Akyaka
Nano-electronics symposium", where this work has been partially
conducted.
%\bibliography{zitate}
\bibliographystyle{elsarticle-num}
%\bibliography{cite}
%\bibliography{siddiki}
%\bibliographystyle{prsty}

\end{document}